\documentstyle[manuscript,eqsecnum,aps]{revtex}
\begin{document}


\preprint{HEP-00-13}

\title{Conjecture on the Interlacing of Zeros in Complex Sturm-Liouville
Problems}

\author{Carl M. Bender${}^1$\cite{bye1}, Stefan Boettcher${}^2$\cite{bye2},
and Van M. Savage${}^1$\cite{bye3}}

\address{${}^1$Department of Physics, Washington University, St. Louis, MO
63130, USA}

\address{${}^2$Department of Physics, Emory University, Atlanta, GA 30322, USA}

\date{\today}

\maketitle

\begin{abstract}
The zeros of the eigenfunctions of self-adjoint Sturm-Liouville eigenvalue
problems interlace. For these problems interlacing is crucial for completeness.
For the complex Sturm-Liouville problem associated with the Schr\"odinger
equation for a non-Hermitian $\cal PT$-symmetric Hamiltonian, completeness and
interlacing of zeros have never been examined. This paper reports a numerical
study of the Sturm-Liouville problems for three complex potentials, the
large-$N$ limit of a $-(ix)^N$ potential, a quasi-exactly-solvable $-x^4$
potential, and an $ix^3$ potential. In all cases the complex zeros of the
eigenfunctions exhibit a similar pattern of interlacing and it is conjectured
that this pattern is universal. Understanding this pattern could provide insight
into whether the eigenfunctions of complex Sturm-Liouville problems form a
complete set.
\end{abstract}


\section{Introduction}
\label{sec1}

The spectra of many classes of non-Hermitian $\cal PT$-symmetric Hamiltonians
are real and positive \cite{R1,R2,R9,R10,R11,R12,S1,S2,S3,S4,S5,S6,S7}. It is
believed that the positivity of the spectra is a consequence of $\cal
PT$ symmetry. Examples of heavily studied ${\cal PT}$-symmetric Hamiltonians
are \cite{R1,R2}
\begin{equation}
H=p^2-(ix)^N\quad(N\geq2)
\label{e1}
\end{equation}
and \cite{R9}
\begin{equation}
H=p^2-x^4+2iax^3+(a^2-2b)x^2+2i(ab-J)x\qquad(J~{\rm integer},~a^2+4b>K_{\rm
critical}),
\label{e2}
\end{equation}
where $K_{\rm critical}$ grows with increasing $J$. For Hamiltonians like those
in (\ref{e1}) and (\ref{e2}) the Schr\"odinger equations for the $k$th
eigenfunction,
\begin{equation}
H\Psi_k(x)=E_k\Psi_k(x),
\label{e3}
\end{equation}
involve a complex potential and may require $x$ to be complex for the boundary
conditions to be defined properly \cite{ROT}. Thus, the Schr\"odinger
eigenvalue problems may be regarded as analytic extensions of Sturm-Liouville
problems into the complex plane.

The eigenfunctions of a conventional self-adjoint Sturm-Liouville problem are
complete. Completeness is the statement that a given function can be represented
as a linear superposition of the eigenfunctions:
\begin{equation}
f(x)=\sum_n a_n \Psi_n(x).
\label{e4}
\end{equation}
It is necessary for the zeros of the eigenfunctions in the complete set to
become dense on the interval in which the Sturm-Liouville problem is defined. If
the zeros did not become dense, it would be impossible to represent a rapidly
varying function \cite{T1,T2}. For conventional Sturm-Liouville problems one can
prove that the zeros of successive eigenfunctions interlace, and this
interlacing of the zeros ensures that the zeros become dense \cite{T1}.

A major open mathematical question for ${\cal PT}$-symmetric Hamiltonians is
whether the eigenfunctions form a complete set. If the zeros of the
eigenfunctions of complex Sturm-Liouville eigenvalue problems exhibit the
property of interlacing, this provides heuristic evidence that the
eigenfunctions might be complete. A proof of completeness would require that we
identify the space in which they are complete, and we do not yet know how to do
this. Nevertheless, if we can understand the distribution of the zeros of the
eigenfunctions, we gain some insight into the question of completeness for
eigenfunctions of $\cal PT$-symmetric Sturm-Liouville problems.

\section{Some Examples of Distributions of Zeros}
\label{sec2}

We have studied three different complex ${\cal PT}$-symmetric Hamiltonians. We
find that in every case the qualitative features of the distribution in the
complex plane of the zeros of the eigenfunctions are very similar: We observe a
shifted interlacing of zeros. We believe that this pattern of zeros is
universal.
\bigskip

\begin{figure}[t]
\vspace{5.0in}
\includegraphics{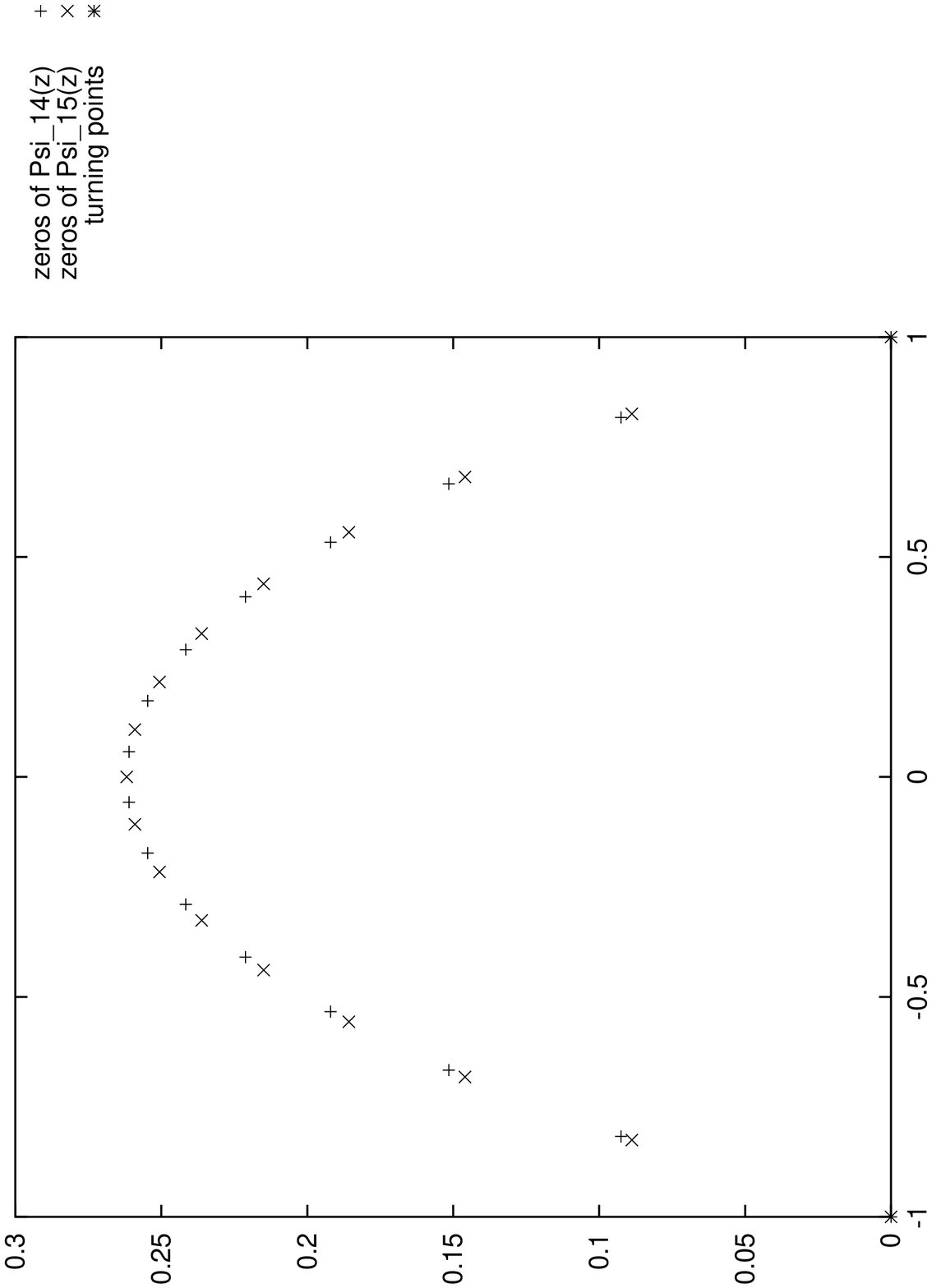}
\caption{Zeros of the $14$th and $15$th eigenfunctions, $\Psi_{14}(x)$ and
$\Psi_{15}(x)$, of the $-(ix)^N$ potential in the large-$N$ limit. The plot
shows the $z$-plane where the turning points have all been scaled to
$-1$ and $1$ as in (2.1). The zeros lie in a small arch-shaped region in
the complex plane. For any two zeros of $\Psi_{15}(z)$, a zero of $\Psi_{14}(z)$
lies between them along the arch-shaped region. This is a complex version of
interlacing.}
\label{f1}
\end{figure}

\noindent{\bf Example 1: Large-$N$ limit of the $-(ix)^N$ potential.} The
large-$N$ limit of a $-(ix)^N$ potential is exactly solvable \cite{R11}. In
Fig.~\ref{f1} the zeros of the $14$th and $15$th eigenfunctions are plotted and
clearly exhibit a form of interlacing in the complex plane. For convenience, we
have scaled the zeros by dividing by the magnitude of the turning points; we
have then performed a linear transformation to fix the turning points at $\pm1$.
The appropriate scaling is
\begin{equation}
z=(xE^{-1/N}+i)N/\pi.
\label{shit}
\end{equation}

\begin{figure}[t]
\vspace{4.0in}
\includegraphics{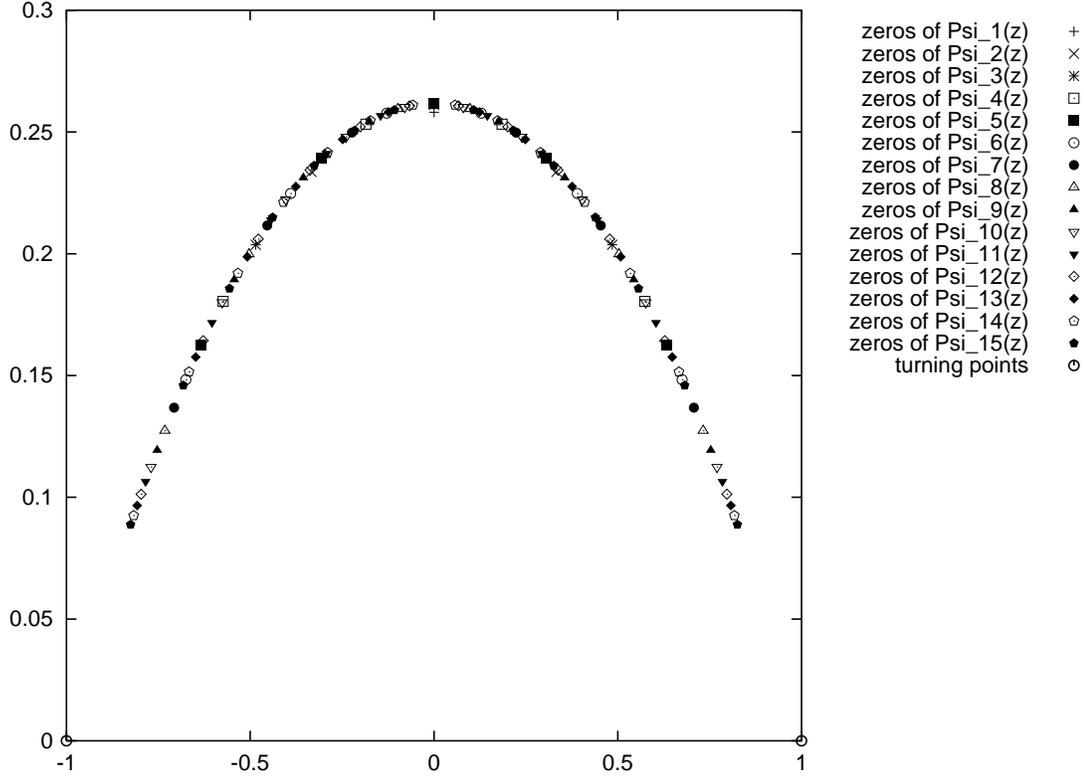}
\caption{Zeros for the first 15 eigenfunctions of the $-(ix)^N$ potential
in the large-$N$ limit. The plot shows the $z$-plane where the turning points
have all been scaled to $-1$ and $1$ as in (2.1). The complex version of
interlacing in Fig.~1 is again evident in this plot, but this plot also
suggests that the zeros are becoming dense in a small region in the complex-$z$
plane.}
\label{f2}
\end{figure}

As shown in Fig.~\ref{f2}, the zeros of the first 15 eigenfunctions interlace
and appear to become dense in a narrow region surrounding an arch-shaped contour
in the complex plane. This contour is the Stokes' line that joins the turning 
points; that is, it is the path along which the phase in the WKB quantization
condition is purely real (and thus the quantum-mechanical wave function is
purely oscillatory)\cite{R1}. (It is interesting that this WKB path differs from
the path that a classical particle follows in the complex plane as it oscillates
between the turning points. The path that a classical particle follows is an
{\it inverted} arch-shaped contour between the same two turning
points.\cite{R2}) 
\bigskip

\begin{figure}[t]
\vspace{4.0in}
\includegraphics{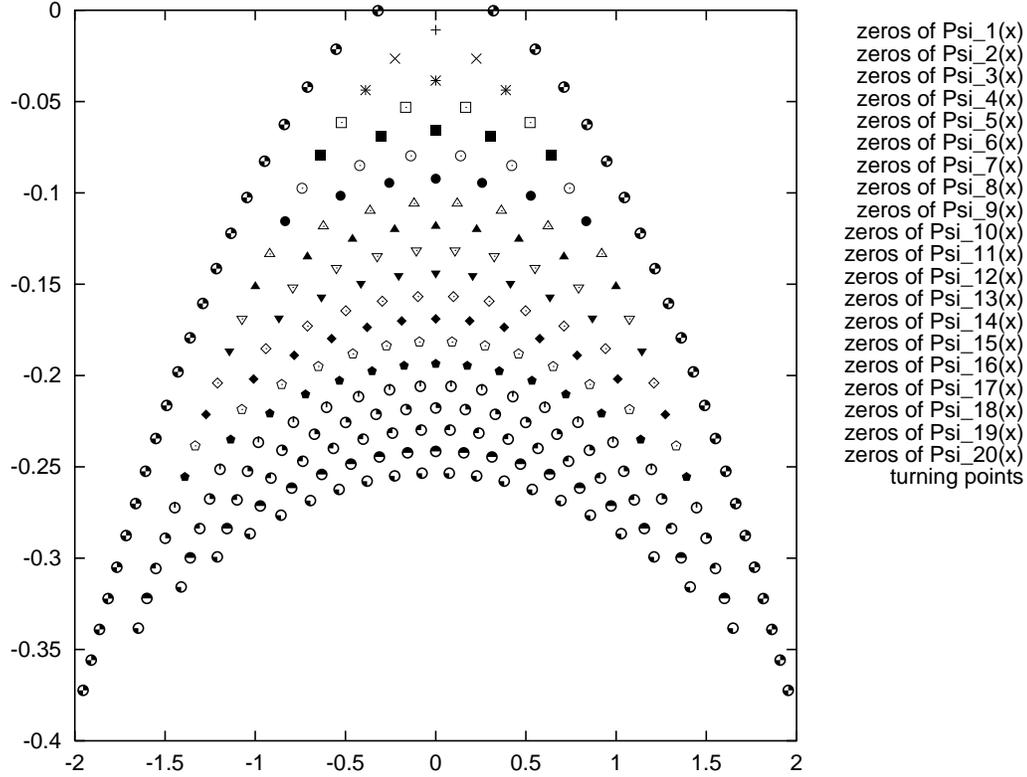}
\caption{Zeros for the 21 exactly solvable eigenfunctions of the
quasi-exactly-solvable $-x^4$ potential in (1.2). The zeros lie in a small 
arch-shaped region in the complex-$x$ plane and exhibit a complex version of 
interlacing. The zeros do not become dense in a region of the complex-$x$ plane 
because the zeros lie on the Stokes' line of the wave function (the curve
along which the wave function is oscillatory). These curves are different for
each eigenfunction; the curves move downward as the energy increases and
there is no limiting curve.}
\label{f3}
\end{figure}

\noindent{\bf Example 2: Quasi-exactly-solvable $-x^4$ potential.} Next,
consider the quasi-exactly-solvable potential in (\ref{e2}) with $a=10$ and
$b=2$. The eigenfunctions of the Schr\"odinger equation have the form of an
exponential multiplied by a polynomial. The zeros of this polynomial are easy to
calculate numerically. For a given $J$ the polynomials in the eigenfunctions all
have the same degree and, as a result, all of the eigenfunctions have the same
number of zeros. However, for the $k$th wave function, $J-k$ zeros lie along the
branch cut on the positive imaginary axis, and we consider these zeros to be 
irrelevant. For values of $J$ ranging from $1$ to $21$, the qualitative behavior
is always the same. In Fig.~\ref{f3} the results for $J=21$ are plotted and the
relevant zeros again lie along the WKB paths in the complex plane. In this case
the zeros are not contained in as narrow a region of the complex plane as for
the $-(ix)^N$ potential because the zeros have not been scaled as in
Fig.~\ref{f2}. The zeros have an imaginary part that becomes more negative as
$k$ increases and exhibit the complex version of interlacing.

\begin{figure}[t]
\vspace{4.0in}
\includegraphics{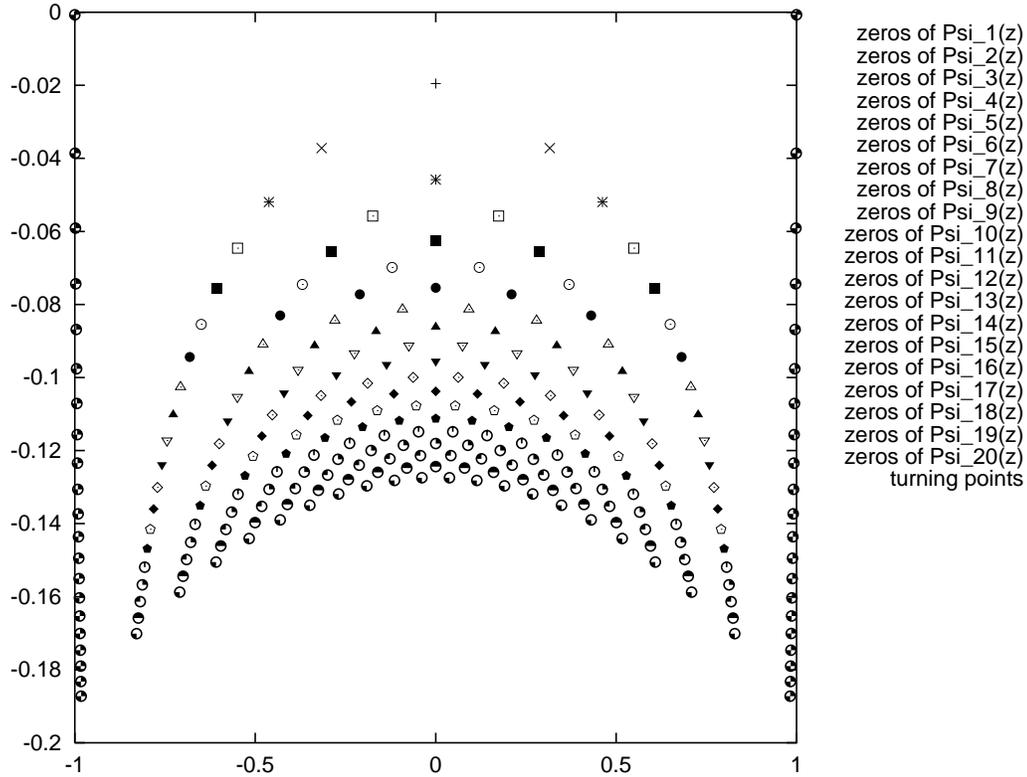}
\caption{Zeros for the 21 exactly solvable eigenfunctions of the
quasi-exactly-solvable $-x^4$ potential (1.2) with the magnitudes of the turning
points fixed. The plot shows the complex-$z$ plane in which the magnitudes 
of the turning points are fixed to unit length. In the $z$ plane 
the zeros lie in a much narrower arch-shaped region than in Fig.~3. Once again,
the zeros exhibit a complex version of interlacing, and now, we believe they
become dense in a narrow region of the scaled complex plane.}
\label{f4}
\end{figure}

In Fig.~\ref{f4} the zeros are scaled so that $z=x/|x_{\rm TP}|$, where
$|x_{\rm TP}|$ are the magnitudes of the classical turning points. (The
classical turning points are the roots of $-x^4+20ix^3+96x^2-2ix=E$.) The
scaled zeros lie in a more compact region in the complex-$z$ plane than the
zeros in Fig.~\ref{f3}. We believe the zeros become dense in this region. Notice
that this arch-shaped region is broader than the corresponding region in
Fig.~\ref{f2} because the turning points do not all lie along the same polar
angle; therefore, the scaling fixes the magnitudes but not the positions of the
turning points. Potentials with various values of $a$ and $b$ were also
investigated and similar results were obtained.
\bigskip

\begin{figure}[t]
\vspace{4.0in}
\includegraphics{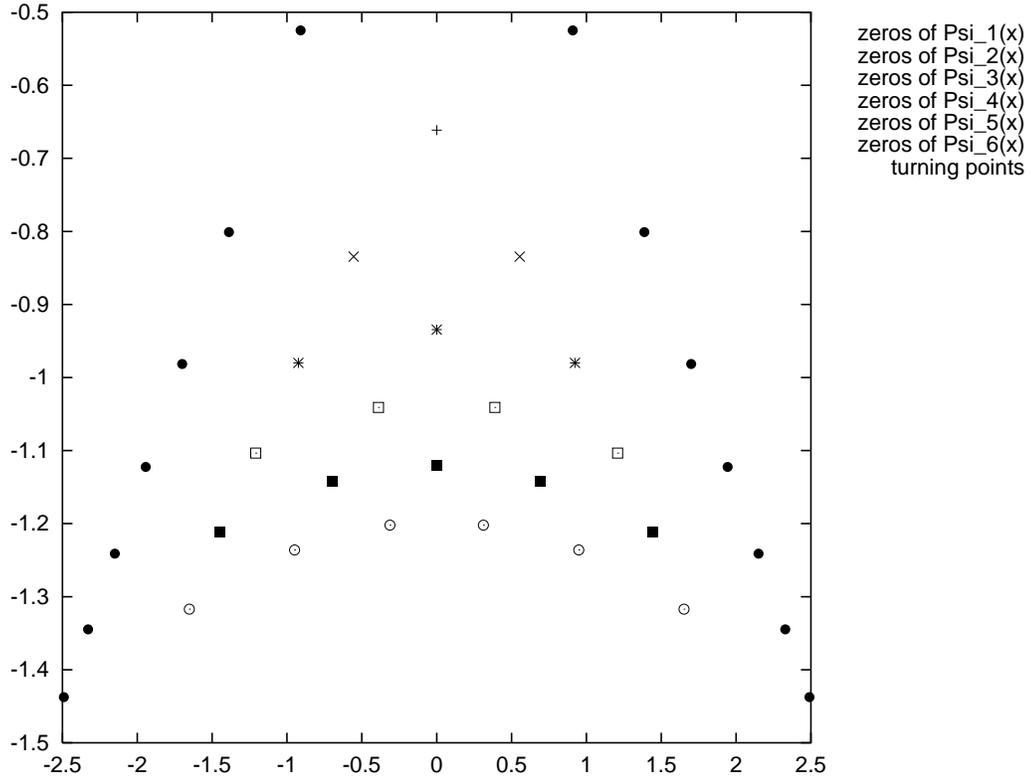}
\caption{Zeros of the first six eigenfunctions of the $ix^3$ potential. These
zeros lie in a small arch-shaped region in the complex plane and exhibit a
complex version of interlacing. The zeros do not become dense in a region of the
complex plane because, as in Fig.~3, the zeros lie on a sequence of curves that
move downward with increasing energy and remain well separated.}
\label{f5}
\end{figure}

\begin{figure}[b]
\vspace{4.0in}
\includegraphics{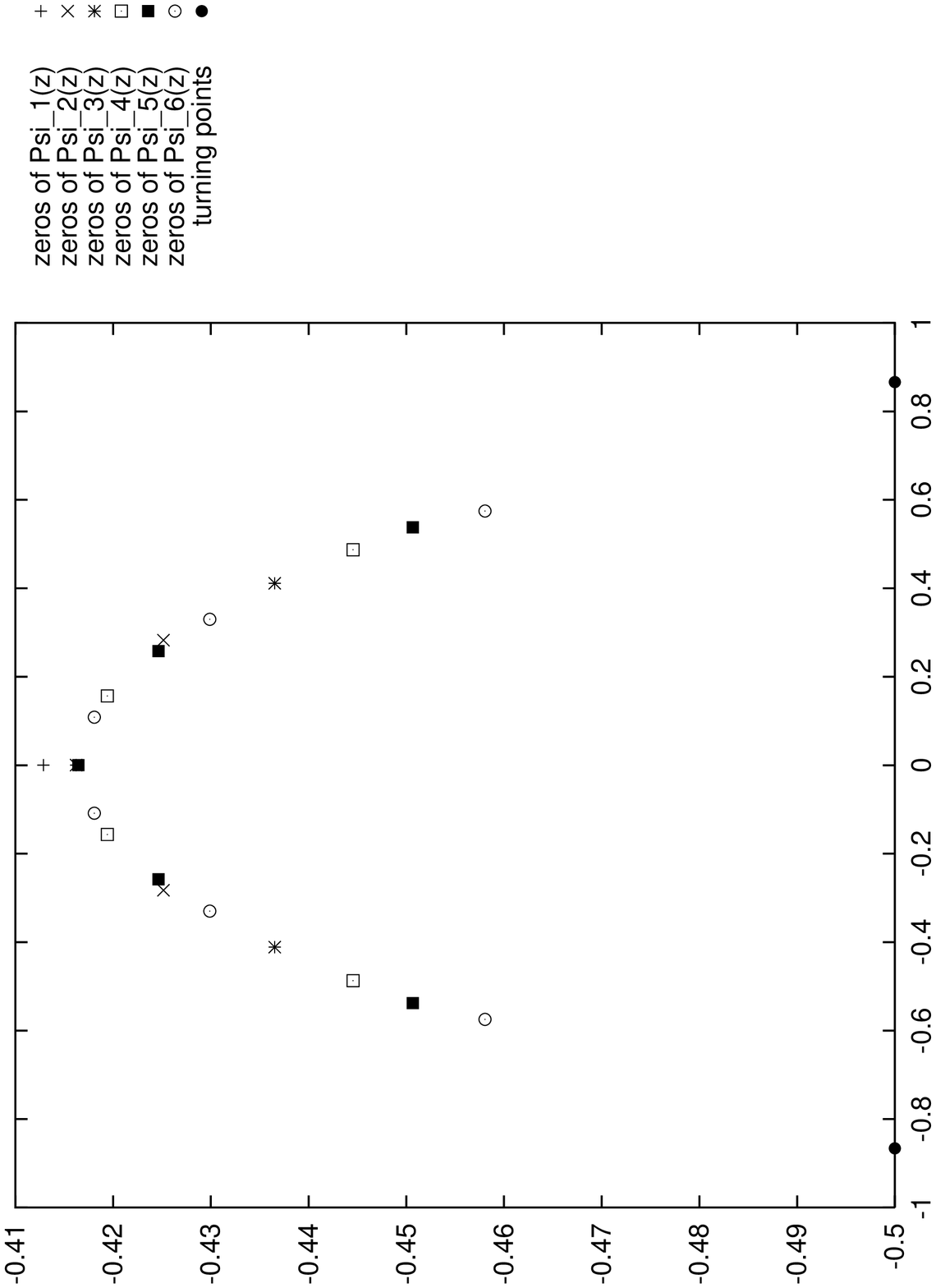}
\caption{Zeros for the first six eigenfunctions of the $ix^3$ potential with the
turning points fixed. The plot is done in the complex-$z$ plane where the
magnitudes of the turning points are fixed to unit length. As a result, the
zeros lie in a narrower arch-shaped region in the complex plane than in Fig.~5.
The zeros exhibit a complex version of interlacing and we believe they become
dense in the scaled complex-$z$ plane.}
\label{f6}
\end{figure}

\noindent{\bf Example 3: $ix^3$ potential} We obtain an $ix^3$ potential when we
set $N=3$ in (\ref{e1}). Using Runge-Kutta techniques in the complex plane, we
have plotted the level curves of the real and imaginary parts of the complex
eigenfunctions. By finding the intersections of these level curves, we have
determined the zeros of the eigenfunctions numerically. These zeros, which are
shown in Fig.~\ref{f5}, lie along the Stokes' lines of the WKB approximation.
Again, the zeros exhibit the complex version of interlacing. They have an
imaginary part that decreases as $k$ increases. In Fig.~\ref{f6} the zeros are
scaled by $z=x/|x_{TP}|=xE^{-1/3}$, which fixes the magnitudes and positions of
the turning points. This plot suggests that after the scaling the zeros become
dense in the complex-$z$ plane. Once again, this plot suggests that the
distribution of zeros in the complex plane is a universal property of complex
Sturm-Liouville eigenvalue problems associated with $\cal PT$-symmetric
Hamiltonians.
\bigskip

\noindent{\bf Statement of the Conjecture}

From our studies we observe that the unscaled zeros of the complex 
eigenfunctions do not become dense on a contour or in a narrow region of the
complex-$x$ plane. In particular, for the $ix^3$ potential WKB theory predicts
that $E_k\sim Ck^{6/5}$ ($k\to\infty$), where $C$ is a constant. Thus, the
turning points behave like $x_{TP}\sim(-iC)^{1/3}k^{2/5}$ ($k\to\infty$) and 
\begin{equation}
{dx_{TP}\over dk}\sim{2\over5}(-iC)^{1/3} k^{-3/5}\quad(k\to\infty).
\label{e9}
\end{equation}
Using Richardson extrapolation we have verified that the distance 
between zeros along the imaginary axis exhibits this $k$-dependence. 
Consequently, the distance from the contour along which the zeros of $\Psi_1$ 
lie to the contour along which the zeros of $\Psi_k$ for large $k$ lie is given
by $\sum_{k=0}^\infty k^{-3/5}$ which is infinitely far away.

We have scaled the zeros by fixing the magnitudes of the turning points relative
to a unit length. After this scaling is performed, the zeros appear to become
dense in a narrow arch-shaped region in the scaled complex plane. If the zeros
do become dense in this narrow region and exhibit the shifted complex version of
interlacing, we conjecture that this behavior suggests that the eigenfunctions
are complete in the scaled complex plane. Since we do not know what space
within which to define completeness, we are unable to give a rigorous proof.

\section*{ACKNOWLEDGMENTS}
\label{s6}
We wish to thank P.~N.~Meisinger for assistance in computer calculations.
We also thank the U.S.~Department of Energy for financial support.


\newpage

\centerline{\bf FIGURE 2}

\vspace{2.0in}

\newpage

\centerline{\bf FIGURE 3}

\vspace{2.0in}

\newpage

\centerline{\bf FIGURE 4}

\vspace{2.0in}

\newpage

\centerline{\bf FIGURE 5}

\vspace{2.0in}

\newpage

\centerline{\bf FIGURE 6}

\vspace{2.0in}


\begin{references}

\bibitem[*]{bye1}E-mail: cmb@howdy.wustl.edu

\bibitem[\dagger]{bye2}E-mail: stb@howdy.wustl.edu

\bibitem[\ddagger]{bye3}E-mail: vmsavage@hbar.wustl.edu

\bibitem{R1} C.~M.~Bender and S.~Boettcher, Phys.~Rev.~Lett.~{\bf 80}, 5243
(1998).

\bibitem{R2} C.~M.~Bender, S.~Boettcher, and P.~N.~Meisinger, J.~Math.~Phys.,
{\bf 40}: 2201, (1999).

\bibitem{R9} C.~M.~Bender and S.~Boettcher, J.~Phys.~A: Math. Gen.~{\bf 31},
L273 (1998).

\bibitem{R10} C.~M.~Bender, F.~Cooper, P.~N.~Meisinger, and V.~M.~Savage,
Phys.~Lett.~A, {\bf 259}, 224-231, (1999).

\bibitem{R11} C.~M.~Bender, S.~Boettcher, H.~F.~Jones, and V.~M.~Savage,
J.~Phys.~A: Math.~Gen., {\bf 32}, 1-11 (1999).

\bibitem{R12} C.~M.~Bender, G.~V.~Dunne and P.~N.~Meisinger, Phys.~Lett.~A,
{\bf 252}, 272 (1999).

\bibitem{S1} A.~A.~Andrianov, M.~V.~Ioffe, F.~Cannata, J.-P.~Dedonder,
Int.~J.~Mod.~Phys.~A {\bf 14}, 2675 (1999).

\bibitem{S2} E.~Delabaere and F.~Pham, Phys.~Lett.~A {\bf 250}, 25 (1998)
and {\bf 250}, 29 (1998).

\bibitem{S3} E.~Delabaere and D.~T.~Trinh, ``Spectral Analysis of the
Complex Cubic Oscillator,'' University of Nice preprint, 1999.

\bibitem{S4} M.~Znojil, Phys.~Lett.~A {\bf 259}, 220 (1999). ???

\bibitem{S5} B.~Bagchi and R.~Roychoudhury, ``A New ${\cal PT}$-Symmetric
Complex Hamiltonian with a Real Spectrum,'' University of Calcutta preprint,
1999.

\bibitem{S6} F.~Cannata, G.~Junker, J.~Trost, Phys.~Lett.~A {\bf 246}, 219
(1998).

\bibitem{S7} H.~Jones, ``The Energy Spectrum of Complex Periodic Potentials of
the Kronig-Penney Type,'' Imperial College preprint, 1999.

\bibitem{ROT} A detailed discussion of the rotation of contours for eigenvalue
problems is given in C.~M.~Bender and A.~Turbiner, Phys.~Lett.~A {\bf 173}, 442
(1993).








\bibitem{T1} E.~L.~Ince, {\it Ordinary Differential Equations} 
(Dover, New York, 1944), Chap.~10.

\bibitem{T2} R.~Courant and D.~Hilbert, {\it Methods of Mathematical Physics} 
(Wiley, New York, 1953), Chap.~6.
\end{references}
\end{document}